\documentclass[runningheads]{llncs}

\usepackage[T1]{fontenc}
\usepackage{graphicx}
\usepackage{array}
\usepackage{url}
\usepackage{tikz}
\usepackage{placeins}
\usetikzlibrary{positioning,arrows.meta,fit,backgrounds}

\graphicspath{{figures/}}

\newcolumntype{L}[1]{>{\raggedright\arraybackslash}p{#1}}
\tikzset{
  cli/.style={draw,thick,fill=blue!8,rounded corners=1pt,
              minimum width=1.4cm,minimum height=0.45cm,
              font=\footnotesize\ttfamily,inner sep=1pt},
  stage/.style={draw,thick,rounded corners=3pt,fill=green!8,
                minimum width=2.05cm,minimum height=0.6cm,
                font=\footnotesize,align=center,inner sep=1pt},
  artifact/.style={draw,thick,fill=yellow!12,
                   minimum width=2.05cm,minimum height=0.5cm,
                   font=\scriptsize,align=center,inner sep=1pt},
  bridge/.style={draw,thick,dashed,fill=gray!8,rounded corners=2pt,
                 font=\scriptsize\itshape,inner sep=2pt,align=center},
  layerlabel/.style={font=\footnotesize\itshape,text=black!65},
  arr/.style={-{Stealth[length=4pt]},semithick},
  arrbridge/.style={-{Stealth[length=3pt]},dashed,semithick,gray}
}

\begin{document}
\title{MARGoS: A Configuration-Driven Platform for Reproducible MARL Experiments in ARGoS}
\titlerunning{MARGoS: Configuration-Driven MARL Experiments in ARGoS}

\author{Tom Philipp Backert\inst{1} \and Mohammad D. Soorati\inst{2} \and Javad Ghofrani\inst{3}}
\authorrunning{T. P. Backert et al.}

\institute{University of L\"ubeck, L\"ubeck, Germany\\
\email{tom.backert@student.uni-luebeck.de}
\and
University of Southampton, Southampton, UK\\
\email{m.soorati@soton.ac.uk}
\and
Bonn-Rhein-Sieg University of Applied Sciences, Sankt Augustin, Germany\\
\email{javad.ghofrani@h-brs.de}}

\maketitle
\begingroup\renewcommand\thefootnote{}\footnotetext{This preprint has not undergone peer review (when applicable) or any post-submission improvements or corrections. The Version of Record of this contribution is published in \emph{Software and Systems Reuse, Product Lines, and Configuration -- VARIABILITY 2026 Satellite Events} (LNCS~17104), and is available online at \url{https://doi.org/10.1007/978-3-032-40421-3_24}.}\endgroup

\begin{abstract}
\sloppy
Multi-agent reinforcement learning (MARL) has become a prominent research direction, particularly within collective robotics, where the ARGoS simulator is widely used to study emergent swarm behavior. Yet despite this growth, systematic support for reuse and reproducibility in MARL-based swarm robotics remains limited and insufficiently standardized: a single experiment is assembled from manually aligned simulator files, training scripts, configuration values, paths, seeds, and environment details, so reuse and reproduction stay fragile even within one research group. To address this gap, we introduce \mbox{MARGoS}, a lightweight, configuration-driven platform developed in this project that elevates these experiment-defining inputs to first-class, platform-managed artifacts. MARGoS enables reproducible experimentation through configuration hashing, end-to-end seed propagation, frozen (immutable) configurations, structured logging, environment fingerprinting, and exportable experiment bundles that facilitate sharing, replay, and verification of results across machines and over time. A five-dimension evaluation (architecture, efficiency, reproducibility, usability, and collaboration) supports our claim that managing experiment variation as a platform responsibility makes ARGoS+MARL experiments easier to reuse, reproduce, and hand off. Our code is available at \url{https://github.com/tombackert/margos}.
\keywords{Variability \and Configuration \and Reproducibility \and Multi-agent reinforcement learning \and Swarm robotics \and ARGoS}
\end{abstract}

\section{Project Context, Goals, and Scope}\label{sec:context_goals_scope}

Multi-agent reinforcement learning (MARL) is a prominent paradigm for training collective robot behavior, in which large numbers of simple, embodied agents produce group-level behavior through decentralized local interaction rather than through a central controller~\cite{sahin2005swarmrobotics,gronauer2022multiagent}. Because experimenting with physical swarms is costly and slow, this training is simulation-first, and ARGoS is a widely used modular simulator for exactly this setting~\cite{pinciroli2012argos,blais2023reinforcement}. Running MARL training on ARGoS, which we refer to as ARGoS+MARL, is therefore a natural and increasingly common way to study learned swarm behavior. Its experiment workflow, however, is largely manual: a single run is assembled by hand from separately edited scenario files, training code, parameters, seeds, output paths, and environment assumptions, so small drifts can silently redefine the experiment and reuse or reproduction stays fragile even within one research group. We present \mbox{MARGoS}, a small, configuration-driven platform built in this project to make reuse and reproduction of ARGoS+MARL experiments a managed property of the platform rather than a matter of user discipline.

The recurring inputs above are variant-bearing experiment artifacts: some define a run, while others capture the state needed to rerun and compare it. When they vary independently, configuration, reproduction, and handoff become manual coordination tasks rather than managed properties. MARGoS treats these experiment-defining inputs as first-class, platform-managed artifacts through configuration hashing, end-to-end seed propagation, frozen configurations, structured logging, environment fingerprinting, and exportable experiment bundles. It thereby turns unmanaged artifact variation into platform-owned managed variation and frames ARGoS+MARL experimentation as a configuration and reuse problem rather than an algorithmic one: the contribution is the platform and its integration of variation-control responsibilities, not a new learning algorithm or swarm behavior.

The gap this addresses is deliberately bounded. Among the adjacent platforms surveyed in Section~\ref{sec:platform_overview}, none is both ARGoS-native and covers configuration $\,\to\,$ orchestration $\,\to\,$ logging $\,\to\,$ analysis $\,\to\,$ export and import as a single pipeline. MARGoS explores integrated experiment-lifecycle support for this stack by centralizing the variant-bearing artifacts, and it was completed as a small-scale university research project sized for meaningful evaluation rather than as a general-purpose product: enough infrastructure to measure workflow efficiency, self-reproducibility, analytical usability, and handoff. Section~\ref{sec:platform_overview} positions the platform against those adjacent systems, Section~\ref{sec:evaluation_design} defines the evaluation it enables, and Section~\ref{sec:results_lessons} reports the lessons it produced.

To show that this problem is real and that MARGoS addresses it correctly, we evaluate the platform along the five dimensions summarized in Table~\ref{tab:goals}, which states each guiding question and the evidence type. RQ1 is enablement-only, validated by whether the platform executes the RQ2-RQ5 evaluation protocols, while RQ2-RQ5 are the four empirically tested dimensions that recur in the evaluation design (Section~\ref{sec:evaluation_design}) and the consolidated results (Section~\ref{sec:results_lessons}). Throughout, the evaluation is supporting evidence for the platform contribution, not the contribution itself.

\begin{table}[t]
\centering
\caption{Project goals and evaluation dimensions. Each dimension maps to a guiding question and to the evidence used to answer it in this paper; RQ1 is enablement-validated through protocol execution, RQ2-RQ5 are threshold-tested.}
\label{tab:goals}
{\small
\setlength{\tabcolsep}{3pt}
\renewcommand{\arraystretch}{1.08}
\begin{tabular}{@{}L{0.23\linewidth}L{0.39\linewidth}L{0.32\linewidth}@{}}
\hline
\textbf{Dimension} & \textbf{Question} & \textbf{Evidence used in this paper} \\
\hline
RQ1 Architecture (enablement) & Can the platform execute the evaluation protocols? & Protocol enablement across RQ2-RQ5 \\
RQ2 Efficiency & Does integration reduce repeated workflow effort? & Time and step reduction vs manual workflow \\
RQ3 Reproducibility & Can identical experiments be rerun reliably on the same hardware? & Reference run plus 20 automated reproductions \\
RQ4 Usability & Does the design reduce interaction complexity? & Heuristic audit and KLM analysis \\
RQ5 Collaboration & Can experiment bundles be handed off reliably to a prepared collaborator? & Export/import handoff trials \\
\hline
\end{tabular}
}
\end{table}

The scope is deliberately narrow: collaboration means asynchronous bundle handoff between prepared environments, not cloud workspaces, shared state, or live co-editing. The project also excludes algorithmic contributions, unrestricted cross-lab portability, and new experiments for this paper, reusing the executed project evaluation only. Each exclusion keeps the evidence tied to managed variation rather than product breadth. The next section turns the variant-bearing artifacts introduced here into platform-owned managed variation.

\section{Platform Overview}\label{sec:platform_overview}

MARGoS is structured as a five-stage pipeline: configuration $\,\to\,$ orchestration $\,\to\,$ logging $\,\to\,$ analysis $\,\to\,$ export/import. Each stage centralizes one or more of the experiment-defining artifacts named in Section~\ref{sec:context_goals_scope}. The pipeline is the concrete mechanism by which those unmanaged inputs become platform-managed variation: configurations define experiment state, orchestration executes it, logs and reports preserve observed behavior, and export/import turns a run into a handoff bundle.

To position this design, we compare it against representative recent platforms that combine multi-agent or multi-robot learning infrastructure with simulator integration. The comparison uses three criteria tied to the gap claim: whether the system is ARGoS-native, whether it covers the full configuration $\,\to\,$ orchestration $\,\to\,$ logging $\,\to\,$ analysis $\,\to\,$ export/import lifecycle, and whether export/import is part of the reproducibility or handoff path rather than a separate manual step. Five comparators cover adjacent multi-agent or multi-robot slices: distributed many-agent infrastructure on Alchemist~\cite{domini2024reusable}, JAX-vectorized learning on the Robotarium testbed~\cite{jain2025jaxrobotarium}, IsaacSim-based MARL self-play~\cite{ke2025arenasim}, Gazebo/ROS autonomous-driving MARL~\cite{liang2024agents}, and AirSim/UE4 UAV swarms~\cite{lou2023airm}. ORBIT~\cite{mittal2023orbit} and IsaacLab~\cite{nvidia2025isaac} cover integrated single-robot learning. Against the three criteria, the surveyed systems provide useful related-work context but do not cover the complete ARGoS+MARL lifecycle evaluated here; the gap claim opened in Section~\ref{sec:context_goals_scope} is therefore bounded to this comparator set rather than to the literature as a whole. ORBIT and IsaacLab are used only as adjacent feasibility evidence for integrated robot-learning infrastructure, while the MARGoS pipeline is derived from the research questions and evaluation metrics for ARGoS+MARL. The contribution claimed here is the RQ-driven integration of variation-control responsibilities for this stack.

The CLI is the user-facing entry point to the pipeline and exposes it as five commands: \texttt{run}, \texttt{show}, \texttt{compare}, \texttt{export}, and \texttt{import}. These commands are entry points into the pipeline rather than one-to-one stage labels: \texttt{run} spans configuration loading, seed propagation, run-state capture, orchestration, and per-iteration logging; \texttt{compare} operates over stored logs and configuration hashes; and \texttt{show}, \texttt{export}, and \texttt{import} operate on stored runs. At the orchestration boundary, ArgosToZoo provides the ARGoS-to-MARL bridge: ARGoS communicates through ZeroMQ with a PettingZoo-compatible Python environment that RLlib training scripts can consume. There is no separate \texttt{reproduce} command: where later metrics label a reproduction task, the platform path is \texttt{import} followed by \texttt{run}, with \texttt{compare} used to verify the rerun against the reference. All five commands are invokable with a single positional argument under a convention-over-configuration default, so the surface a user must traverse to move an experiment through the pipeline is fixed at a five-verb interface. This five-command surface is itself a design decision, and it is the user-facing measurement surface that the efficiency, usability, and collaboration evaluations in Section~\ref{sec:results_lessons} measure against.

Configuration hashing, seed propagation, frozen configs, structured logs, environment fingerprints, and exported source copies are the core reusable assets through which the platform turns experiment variants into managed artifacts. The hash is computed over the normalized frozen configuration. Because that configuration contains file references rather than embedded file bodies, the config hash alone is not a source-file versioning mechanism. Local rerun identity therefore assumes unchanged referenced sources; the platform makes that boundary explicit rather than claiming content-level version control. Drift in referenced simulator files and training scripts is controlled at the bundle boundary: the completed run records the frozen config, integrity hash, environment fingerprint, and logs, while the exported bundle copies the scenario and training-script sources and import rewrites references to those bundled copies. A received bundle therefore reproduces the captured variant rather than resolving mutable paths in the sender's working tree.

\begin{figure}[!t]
\centering
\resizebox{\linewidth}{!}{%
\begin{tikzpicture}[node distance=0.18cm and 0.14cm]
\node[cli] (cli-run)    {run};
\node[cli,right=of cli-run]     (cli-show)    {show};
\node[cli,right=of cli-show]    (cli-compare) {compare};
\node[cli,right=of cli-compare] (cli-export)  {export};
\node[cli,right=of cli-export]  (cli-import)  {import};
\node[layerlabel,left=0.25cm of cli-run] {CLI};

\node[stage,below=0.55cm of cli-run.south west,anchor=north west] (st-config) {configuration};
\node[stage,right=of st-config]   (st-orch)     {orchestration};
\node[stage,right=of st-orch]     (st-log)      {logging};
\node[stage,right=of st-log]      (st-analysis) {analysis};
\node[stage,right=of st-analysis] (st-export)   {export/import};
\node[layerlabel,left=0.25cm of st-config] {Pipeline};

\node[artifact,below=0.45cm of st-config] (art-config) {frozen configs};
\node[artifact,right=of art-config]   (art-runs)   {runs $+$ seeds};
\node[artifact,right=of art-runs]     (art-logs)   {structured logs};
\node[artifact,right=of art-logs]     (art-reports){reports};
\node[artifact,right=of art-reports]  (art-bundles){bundles $+$ fingerprints};
\node[layerlabel,left=0.25cm of art-config,align=right] {Managed\\artifacts};

\draw[arr] (cli-run.south) to[bend right=8] (st-config.north);
\draw[arr] (cli-run.south) -- (st-orch.north);
\draw[arr] (cli-run.south) to[bend left=8] (st-log.north);
\draw[arr] (cli-show)    -- (st-log.north);
\draw[arr] (cli-compare) -- (st-analysis.north);
\draw[arr] (cli-export)  -- (st-export.north);
\draw[arr] (cli-import)  -- (st-export.north);

\draw[arr] (st-config)   -- (st-orch);
\draw[arr] (st-orch)     -- (st-log);
\draw[arr] (st-log)      -- (st-analysis);
\draw[arr] (st-analysis) -- (st-export);

\draw[arr] (st-config.south)   -- (art-config.north);
\draw[arr] (st-orch.south)     -- (art-runs.north);
\draw[arr] (st-log.south)      -- (art-logs.north);
\draw[arr] (st-analysis.south) -- (art-reports.north);
\draw[arr] (st-export.south)   -- (art-bundles.north);

\node[bridge,below=0.35cm of art-runs] (bridge) {ArgosToZoo bridge\\(ARGoS C\texttt{++} $\leftrightarrow$ ZeroMQ $\leftrightarrow$ Python/RLlib)};
\draw[arrbridge] (st-orch.south west) to[bend right=20] (bridge.north west);
\draw[arrbridge] (bridge.north east) to[bend right=20] (st-orch.south east);
\end{tikzpicture}%
}
\caption{MARGoS architecture. Control flow runs top-down from the five-command CLI through the five-stage pipeline (configuration $\to$ orchestration $\to$ logging $\to$ analysis $\to$ export/import); managed-artifact flow runs from each stage into the bottom row (frozen configs, runs and seeds, structured logs, reports, exportable bundles with environment fingerprints). The \texttt{run} command intentionally fans out because it spans configuration, orchestration, and logging rather than naming one stage. The ArgosToZoo bridge couples the orchestration stage to ARGoS via ZeroMQ.}
\label{fig:architecture}
\end{figure}
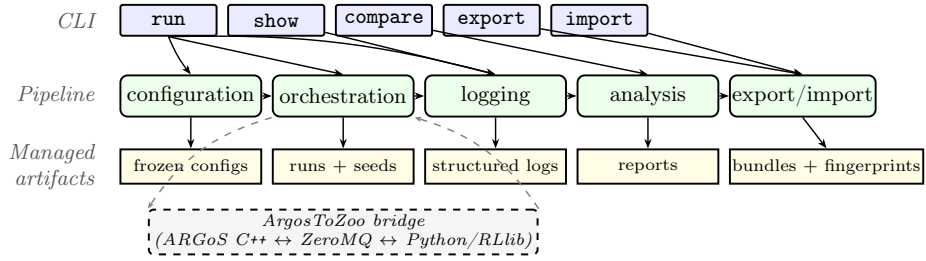

The asset-to-outcome mapping is deliberately narrow. Configuration hashing, seed propagation, and frozen configs support RQ3 same-hardware rerun identity; logs and summaries support RQ2 workflow-effort measurement and result inspection; the RQ4 analytical-usability claim comes primarily from the five-command surface and measured task paths rather than from logs alone; and export/import bundles assembled from frozen configs, logs, source copies, and fingerprints support the RQ5 handoff evaluation. Figure~\ref{fig:architecture} summarizes this control and managed-artifact flow across the five pipeline stages, and Section~\ref{sec:evaluation_design} picks up the asset-to-outcome mapping as the input to the evaluation design.

\FloatBarrier

\section{Evaluation Design}\label{sec:evaluation_design}

Section~\ref{sec:platform_overview} mapped each platform-owned asset to an outcome it enables; this section defines how those outcomes are measured. Evaluation was predefined, metric-driven, and based on counted, timed, or analytical measures: metrics, thresholds, and sample sizes were frozen before data collection, and no new experiments were added for this paper. RQ2, RQ4, and RQ5 compare documented manual workflows against MARGoS paths; RQ3 reruns one same-hardware reference run. In the timed protocols, trial order was documented and conditions were interleaved where the protocol allowed, so learning effects could be checked rather than inferred away. Surveys, interviews, and participant self-report ratings were excluded; RQ4 uses Nielsen-style heuristic evaluation and GOMS/KLM modelling~\cite{nielsen1990heuristic,molich1990improving,card1980keystrokelevel,card1983psychology}, as expert analytical evidence rather than behavioral user evidence. Primary interpretation is threshold/effect-size based; Mann--Whitney and bootstrap checks are reported only as robustness checks for RQ2/RQ5 timing, not as population-level inference.

Table~\ref{tab:evaluation-design} consolidates the four empirical dimensions. It also fixes the operational readings needed to assess fairness: RQ2 uses screen-recorded human-effort time and discrete workflow steps; RQ3 requires reward and AUC within $\pm$1\% with matching configuration identity; RQ4 combines 35 heuristic criteria with weighted KLM task paths; and RQ5 includes step counts and timing, with handoff success defined as final-50-reward match within $\pm$1\%. RQ1 Architecture is answered separately by whether the architecture enables the RQ2-RQ5 protocols to run end-to-end.

\begin{table}[t]
\centering
\caption{Executed evaluation design: primary metrics, sample sizes, and predefined practical targets across the four empirically tested dimensions. RQ1 Architecture is omitted because it is validated by protocol enablement: the platform must execute the RQ2-RQ5 protocols end-to-end rather than pass a separate threshold test.}
\label{tab:evaluation-design}
{\footnotesize
\setlength{\tabcolsep}{3pt}
\renewcommand{\arraystretch}{1.08}
\begin{tabular}{@{}L{0.20\linewidth}L{0.30\linewidth}L{0.30\linewidth}L{0.14\linewidth}@{}}
\hline
\textbf{Dimension} & \textbf{Primary metrics} & \textbf{Sample} & \textbf{Target} \\
\hline
RQ2 Efficiency & Time-to-Complete; Steps-to-Complete & N=5 per condition & $\geq$50\% reduction \\
RQ3 Reproducibility & Reproduce-Success-Rate; Result-Variance & N=20 automated runs & $\geq$90\% \\
RQ4 Usability & Heuristic compliance; KLM reduction & 35 criteria; 7 tasks; 1 structured audit & $\geq$80\%; $\geq$50\% \\
RQ5 Collaboration & Steps/time-to-share; first-run/reproduce time; handoff success & 10 manual + 10 platform cycles; timed N=10/condition & $\geq$50\%; $\geq$90\% success \\
\hline
\end{tabular}
}
\end{table}
\FloatBarrier

Four measurement boundaries scope the Section~\ref{sec:results_lessons} results. RQ3 is same-hardware only. RQ5 is local-first and asynchronous, but the measured proxy is narrower: a prepared local collaborator account with isolated home directory, fixed repository copies, one shared preinstalled RQ5 virtualenv, and transfer-only artifact access; network and machine provisioning are excluded. Human-effort timing excludes compute wait. Behavioral usability metrics (Time-to-First-Success, Error-Rate, Recovery-Time, Documentation-Lookups) were dropped because the self-as-user setup invalidates behavioral interpretation, so RQ4 reports only analytical methods. The next section reports the measured outcomes against these targets and boundaries as the evidence bundle that anchors the paper's contribution.

\section{Results and Lessons Learned}\label{sec:results_lessons}

\begin{figure}[t]
\centering
\includegraphics[width=\linewidth]{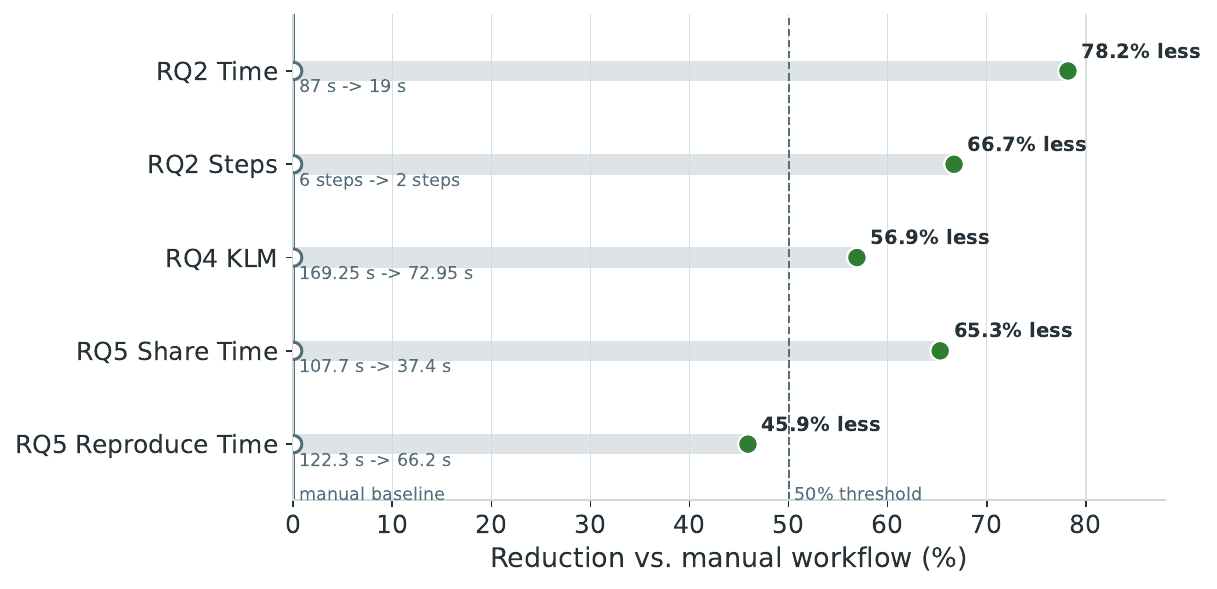}
\caption{Cross-dimension effort reductions on the executed project evaluation: every metric is plotted against the manual baseline at 0\%, and the predefined $\geq$50\% target appears as a dashed line. The chart is a visual comparison, not pooled inference: it juxtaposes workflow-effort gains on the RQ2 reference task, analytical KLM reductions on the RQ4 task set, and sharing-phase gains on RQ5, while showing the RQ5 reproduce-time metric narrowly below the 50\% target.}
\label{fig:cross-srq}
\end{figure}

Table~\ref{tab:results} summarizes the empirical outcomes: RQ2-RQ4 meet their predefined practical targets, while RQ5 is partially supported because two timing targets fall short despite reliable handoff. Fig.~\ref{fig:cross-srq} places the cross-RQ effort-reduction metrics on a single axis, with the dashed line marking the predefined $\geq$50\% target. The four lessons below unpack what each judgment means at the project level; together they form the evidence bundle that anchors the configuration-driven contribution introduced in Section~\ref{sec:context_goals_scope}.

\begin{table}[!ht]
\centering
\caption{Measured outcomes across the four empirically tested dimensions. RQ2-RQ4 meet their predefined practical targets, while RQ5 is partially supported; jointly the four rows are the evidence bundle that platform-owned managed variation translates into measurable workflow effort reduction, deterministic self-reproducibility, analytical usability gains, and verified handoff at a first-run overhead.}
\label{tab:results}
{\footnotesize
\setlength{\tabcolsep}{3pt}
\renewcommand{\arraystretch}{1.08}
\begin{tabular}{@{}L{0.20\linewidth}L{0.36\linewidth}L{0.20\linewidth}L{0.16\linewidth}@{}}
\hline
\textbf{Dimension} & \textbf{Metric result} & \textbf{Threshold} & \textbf{Judgment} \\
\hline
RQ2 Efficiency & Time -78.2\%; steps -66.7\% & $\geq$50\% reduction & Supported \\
RQ3 Reproducibility & 100\% success (20/20); 0.0000 SD & $\geq$90\% success & Strongly\newline supported \\
RQ4 Usability & 85.7\% heuristic compliance; 56.9\% KLM reduction & $\geq$80\%; $\geq$50\% & Supported \\
RQ5 Collaboration & 100\% handoff success; share time -65.3\%; reproduce time -45.9\%; first run +7.4\% slower & $\geq$50\% reductions; $\geq$90\% success & Partially supported \\
\hline
\end{tabular}
}
\end{table}
\FloatBarrier

Time-to-Complete on the RQ2 reference task dropped from $87.0 \pm 15.02$~s in the manual condition to $19.0 \pm 1.41$~s in the platform condition, a 78.2\% reduction, and Steps-to-Complete dropped from 6.0 to 2.0, a 66.7\% reduction; both exceeded the predefined $\geq$50\% threshold (robustness: $U{=}0$, exact $p{=}0.011$, bootstrap relative-reduction CI $[74.4,80.9]$\%). The gains concentrated in operational coordination, including training initiation, monitoring, result extraction, and export, rather than in the underlying robotics or learning algorithms, because the platform replaces a multi-tool, multi-step manual sequence with single-command operations. The project lesson is that integration gives the clearest workflow benefit when it removes repeated manual coordination across tools, not when it adds features. This is a measured reduction on the RQ2 reference task with $N{=}5$ per condition and a single expert evaluator, not a generalized productivity claim.

All 20 RQ3 reruns matched the reference on final reward and area under the learning curve (100\% success, 0.0000 SD), and the \mbox{Config-Integrity} and \mbox{Seed-Determinism} checks passed; this exceeds the predefined $\geq$90\% threshold. This is the clearest single-dimension result in the bundle. Within the same-hardware boundary stated in Section~\ref{sec:evaluation_design}, the platform's centralized configuration management and deterministic seed propagation produced perfectly reproducible training outcomes. The project lesson is that bounded self-reproducibility follows directly from making configuration identity, seed propagation, and runtime configuration-integrity checks platform responsibilities rather than relying on user discipline to manually track them; structured logs support auditability of these checks but are not the directly measured mechanism here. Within the evidence bundle this is the rerun-identity strand: managed configuration and seeds preserve rerun identity, while the coordination strand (RQ2) and the handoff strand (RQ5) carry their own outcomes below.
\FloatBarrier

Heuristic-Compliance-Rate was 30/35 (85.7\%) against a $\geq$80\% threshold, and KLM-Reduction was 56.9\% weighted across seven tasks against a $\geq$50\% threshold; both targets were met. The KLM reduction is not uniform across the task set: gains concentrated in execution, monitoring, results access, export, and import/run reproduction tasks (T3-T7, 58-88\% reductions), while configuration editing tasks (T1-T2) showed 0\% reduction because both conditions use the same editor for the same parameters. The project lesson is that analytical usability improves where the platform owns the task path; it does not improve where the user still hand-edits configuration files. The 0\% reduction on T1-T2 is a genuine design boundary rather than a measurement artifact, and it identifies the most actionable area for future platform work.

\begin{figure}[!htb]
\centering
\includegraphics[width=\linewidth]{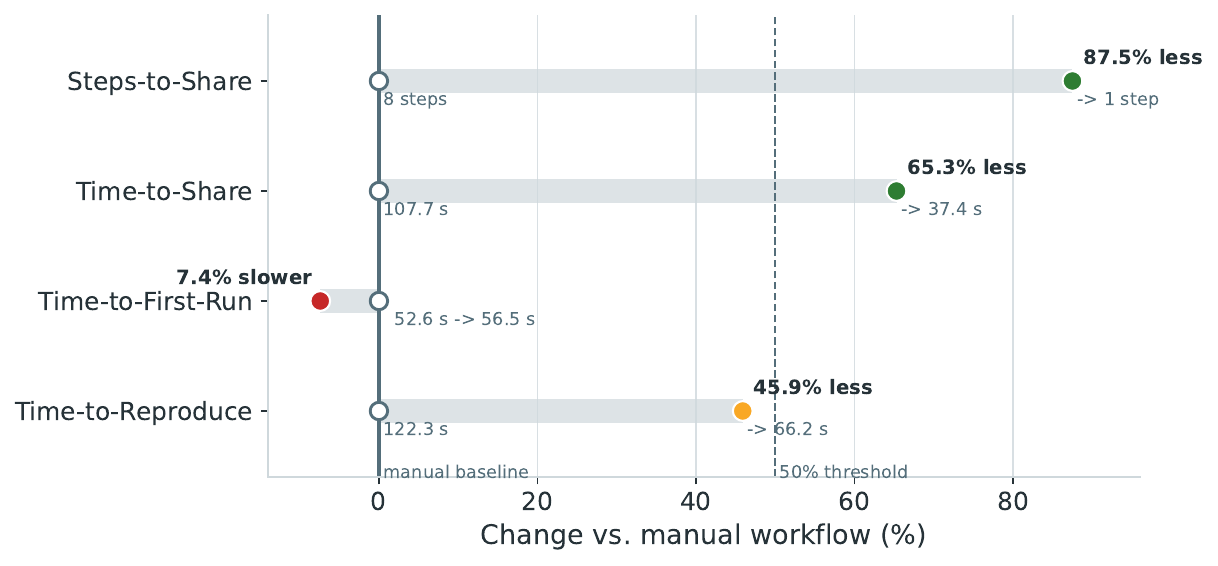}
\caption{Collaboration tradeoff (RQ5): the four timing metrics relative to the manual baseline at 0\%. Large sharing-phase gains and reliable handoff success are set against the first-run overhead introduced by bundle validation and import checks. The figure shows the RQ5 partial verdict as a single tradeoff in the evidence bundle rather than as a missed target in isolation.}
\label{fig:srq5}
\end{figure}

\begin{sloppypar}
Handoff-Success-Rate was 100\% in both conditions (10/10 platform \mbox{handoffs}, 10/10 manual baseline); Steps-to-Share dropped from 8 to 1 ($-$87.5\%); \mbox{Time-to-Share} dropped from $107.7 \pm 24.1$~s to $37.4 \pm 18.2$~s ($-$65.3\%; robustness: $U{=}0$, exact $p{<}0.001$, bootstrap relative-reduction CI $[53.8,75.7]$\%); \mbox{Time-to-Reproduce} dropped from $122.3 \pm 30.3$~s to $66.2 \pm 13.1$~s ($-$45.9\%), narrowly below the predefined 50\% threshold; and \mbox{Time-to-First-Run} was 7.4\% slower in the platform condition ($52.6 \pm 11.4$~s vs $56.5 \pm 11.6$~s). Because both conditions reached 100\% success, RQ5 differentiates effort and overhead rather than a reliability gain; Bundle-Completeness was 8/8, and Setup-Divergence averaged 3.0 package mismatches that did not prevent reproduction. Figure~\ref{fig:srq5} visualizes this tradeoff: large sharing-phase gains and reliable success, set against the first-run overhead. The recorded handoff timings identify the recurring overhead in the platform's import processing: bundle extraction, configuration path rewriting, and environment fingerprint comparison. The project lesson is that verified handoff preserves 100\% handoff success while reducing sharing effort, at a first-run overhead from bundle validation and import checks; RQ5 is partially supported, and this tradeoff is the project's most actionable finding for managed-variant infrastructure because it surfaces a real cost of that infrastructure. This leads directly to the validity and transferability boundaries in Section~\ref{sec:limitations_transferability}.
\end{sloppypar}

\FloatBarrier

\section{Limitations and Transferability}\label{sec:limitations_transferability}

The four-validity-type framework of internal, construct, external, and conclusion validity is the standard taxonomy for validity threats in empirical software-engineering studies and organizes the boundaries under which the Section~\ref{sec:results_lessons} results hold~\cite{cook1979quasiexperimentation,wohlin2012experimentation}. The core internal-validity limit is single developer-evaluator, self-evaluation: instrumented metrics reduce evaluator discretion for RQ2, RQ3, and RQ5, while RQ4 remains judgment-heavy and is bounded by documented heuristic-audit and KLM decision rules (Section~\ref{sec:evaluation_design}). The core construct-validity limit is that KLM predicts expert error-free interaction time and a heuristic audit is not novice-user behavior, so the RQ4 claim is a design-property claim, not evidence of novice behavior, real-user error patterns, or real-user time savings. The core external-validity limits are the measurement boundaries stated in Section~\ref{sec:evaluation_design}: same-hardware self-reproducibility, a prepared collaborator account despite measured residual library drift, and the ARGoS/RLlib stack and executed task set as the only evaluated context. The conclusion-validity limit is that results are threshold and effect-size judgments under the executed study design, not population-level inference; predefined metrics and thresholds reduce post-hoc selection risk, documented trial order addresses learning effects in the timed protocols, and RQ5's import-overhead isolation addresses the main remaining handoff confound.

Within those boundaries, treating explicit configuration, deterministic seed propagation, structured logs, and exportable bundles as first-class managed-variation mechanisms is the reusable design decomposition this project advances; it is a candidate pattern for simulator-learning stacks in which those primitives can be exposed, not an empirically validated transfer claim. Direct portability is not claimed without further evidence: transferability here is a pattern claim, not a deployment claim.

\section{Conclusion}\label{sec:conclusion}

MARGoS, a configuration-driven platform for ARGoS+MARL experimentation, turns fragmented experiment-defining artifacts into an executable, reusable pipeline. Five research questions evaluate it: the platform answers RQ1 by running the RQ2-RQ5 protocols end-to-end. The measured pattern is sharp but bounded: integration cuts RQ2 workflow time by 78.2\%, RQ3 reruns reach 100\% success on the same hardware with no observed result variance, RQ4 meets its analytical usability targets, and RQ5 preserves 100\% handoff success with 65.3\% less sharing time while still missing the reproduce-time target and adding a 7.4\% first-run overhead. The resulting next steps are concrete: reduce import-validation overhead and test pattern-level transfer toward cross-lab portability in simulator-learning stacks that expose the same primitives.

{\raggedright
\bibliographystyle{splncs04}
\bibliography{references}
}

\end{document}